\documentclass[10pt,aps,prl,reprint,twocolumn, nofootinbib,aps,superscriptaddress]{revtex4-1}
\usepackage[linkcolor = blue, citecolor = blue, urlcolor = blue, colorlinks = true]{hyperref}
\usepackage[usenames,dvipsnames]{xcolor}
\usepackage[normalem]{ulem}
\usepackage{graphicx}
\usepackage{listings}
\usepackage{stmaryrd}
\usepackage{amssymb}
\usepackage{amsmath}
\usepackage{gensymb}
\usepackage{float}
\usepackage{bbold}
\usepackage{bm}

\newcommand{\St}{\bm{\sigma}}
\newcommand{\Ss}{\bm{\sigma}^{({\rm s})}}
\newcommand{\Sd}{\bm{\sigma}^{({\rm d})}}
\newcommand{\Se}{\bm{\sigma}^{({\rm e})}}
\newcommand{\Sv}{\bm{\sigma}^{({\rm v})}}
\newcommand{\Sr}{\bm{\sigma}^{({\rm r})}}
\newcommand{\traceless}[1]{\left\llbracket #1 \right\rrbracket}

\hypersetup{
    colorlinks=true,
    linkcolor=blue,
    citecolor=red}

\begin{document}

\title{Hydrodynamic enhancement of $p-$atic defect dynamics}

\author{Dimitrios Krommydas}
\affiliation{Instituut-Lorentz, Universiteit Leiden, P.O. Box 9506, 2300 RA Leiden, The Netherlands}
\author{Livio Nicola Carenza}
\affiliation{Instituut-Lorentz, Universiteit Leiden, P.O. Box 9506, 2300 RA Leiden, The Netherlands}
\author{Luca Giomi}
\email{giomi@lorentz.leidenuniv.nl}
\affiliation{Instituut-Lorentz, Universiteit Leiden, P.O. Box 9506, 2300 RA Leiden, The Netherlands}

\date{\today}

\begin{abstract}
We investigate numerically and analytically the effects of hydrodynamics on the dynamics of topological defects in $p-$atic liquid crystals, i.e. two-dimensional liquid crystals with $p-$fold rotational symmetry. Importantly, we find that hydrodynamics fuels a generic passive self-propulsion mechanism for defects of winding number $s=(p-1)/p$ and arbitrary $p$. Strikingly, we discover that hydrodynamics always accelerates the annihilation dynamics of pairs of $\pm 1/p$ defects, and that, contrary to expectations, this effect increases with $p$. Our work paves the way towards understanding cell intercalation and other remodelling events in epithelial layers.
\end{abstract}

\maketitle

The physics of topological defects in liquid crystals have experienced in the last decade a tremendous revival, thanks to a wealth of exciting discoveries at the interface between soft condensed matter and biological physics \cite{Sanchez2012, Keber2014, Giomi2014, Giomi2015, Doostmohammadi2017, Guillamat2018, Doostmohammadi2018, Carenza2019, Carenza2019_2}. The most common class of liquid crystal defects, known as {\em disclinations}, consists of point or line singularities around which the average orientation of the anisotropic building blocks undergoes one or more complete revolutions, thereby disrupting the local orientational order~\cite{Mermin1979, chaikin_lubensky_1995, Giomi2017, Pollard2019, Carenza2022, lopez-leon2011}. In $p-$atic liquid crystals $-$ two-dimensional liquid crystals with $p-$fold rotational symmetry, among which nematics ($p=2$) and hexatics ($p=6$) are the best known examples $-$ defects can be classified in terms of their winding number or strength $s$; that is, the number of revolutions of the orientation field along an arbitrary loop enclosing the defect core, i.e. $s=\pm 1/p,\,\pm 2/p$, etc. \cite{PhysRevLett.82.2721,https://doi.org/10.1002/cphc.200900755,PhysRevLett.118.158001,Zhao2012LocalCS}.

Although the equilibrium physics of liquid crystal defects represents a mature topic across several areas of physics $-$ from cosmology \cite{Kibble_1976}, down to condensed matter~\cite{chaikin_lubensky_1995, Halperin1978} and particle physics~\cite{Dirac1931,HOOFT1974276,Polyakov:1974ek} $-$ our understanding of their dynamics is still in a phase of accelerated expansion, especially in the realm of biological matter, where defects have been suggested to accomplish various vital functions. These include driving the extrusion of apoptotic cells in epithelial layers~\cite{saw2017,Monfared2021}, coordinating large scale cellular flows during wound healing and morphogenetic events~\cite{Brugues2014,kawaguchi2017}, and seeding the development of non-planar features, such as tentacles and protrusion in simple organisms, such as {\em Hydra}~\cite{Livshits2017,Braun2018,Livshits2021,Maroudas-sacks2021}. 

While the biochemical aspects of these processes are mostly understood, less is known about the role of physical interactions. Their origin, nevertheless, can be single-handedly ascribed to the existence of a hydrodynamic phenomenon known as {\em backflow}; the hydrodynamic flow resulting from spatial variations of the average microscopic orientation~\cite{Berreman:1975,doi:10.1080/15421407308083360, microflows, TSUJI2021112386, doi:10.1080/02678290110067236, cryst11040430}. In passive liquid crystals, departure from the  uniformly oriented equilibrium configuration is generally transient and often originates from a sudden change in the environmental conditions, such as the abrupt variation of an external electric or magnetic field in optical devices~\cite{Brochard1972,Tiribocchi2014}. Conversely, in active systems, distortions occur spontaneously as a consequence of the internal stresses collectively exerted by the active subunits~\cite{Sanchez2012, hatwalne2004, Giomi2015}. 

Whether passive or active, backflow significantly affects the static and dynamical behavior of topological defects. In passive nematic liquid crystals, for instance, this effect is known to affect the annihilation dynamics of neutral pairs of elementary disclinations~\cite{toth2002,toth2003hydrodynamics,PhysRevA.46.7765,PhysRevLett.95.097802,PhysRevE.71.061709,PhysRevLett.95.027801,PhysRevE.85.021703,PhysRevResearch.2.013080}. In active nematics $-$ such as {\em in vitro} mixture of cytoskeletal filaments and motor proteins~\cite{Sanchez2012, Keber2014,Martinez-Prat2021, Colen2021} or certain types of prokaryotic~\cite{Dunkel2013,Wensink2012} and eukaryotic~\cite{saw2017,Balasubramaniam2021} cells $-$ backflow drives the propulsion of $s=1/2$ defects and influences the hydrodynamic stability of active layers with respect to non-planar deformations~\cite{Keber2014, Metselaar2019, Hoffmann2022,Alert2022}. Yet, a hydrodynamic theory that captures backflow effects in liquid crystalline systems with generic $p-$atic symmetry was developed only recently~\cite{Giomi:2021a,Giomi:2021b}. Hence, the current understanding of defect dynamics in such systems is still in its infancy. 

In this article we bridge this gap.  Employing the theory of ~\cite{Giomi:2021a,Giomi:2021b} we make \textit{essential} steps towards solving the critical problem of passive $p-$atic defect dynamics, whose understanding is imperative for a complete and consistent description of any $p-$atic liquid crystalline system. First, accounting for arbitrary $p-$fold symmetry, we calculate the velocity field of arbitrary $p-$atic disclinations leveraging on recent progress toward generalizing the classic hydrodynamic theory of hexatic liquid crystals~\cite{Zippelius:1980a,Zippelius:1980b}. Strikingly, we find that backflow fuels a generic passive self-propulsion mechanism for defects of winding number $s=(p-1)/p$ and arbitrary $p$ values. Although this mechanism is not unique to nematics, we find that nematics are the only $p-$atic liquid crystals in which passive self-propulsion is thermodynamically stable. Furthermore, we analyze the effect of hydrodynamics on the annihilation of neutral elementary defect pairs $s = \pm 1/p$. Crucially, we discover that backflow always accelerates their annihilation dynamics, and that, contrary to expectations, becomes increasingly more relevant as $p$ increases. Finally, we uncover that, surprisingly, the source of this acceleration is generically different than that of self-propulsion, which becomes equally important only in the case of nematics. 

\begin{figure}[t!]
\centering
  \includegraphics[width=\columnwidth]{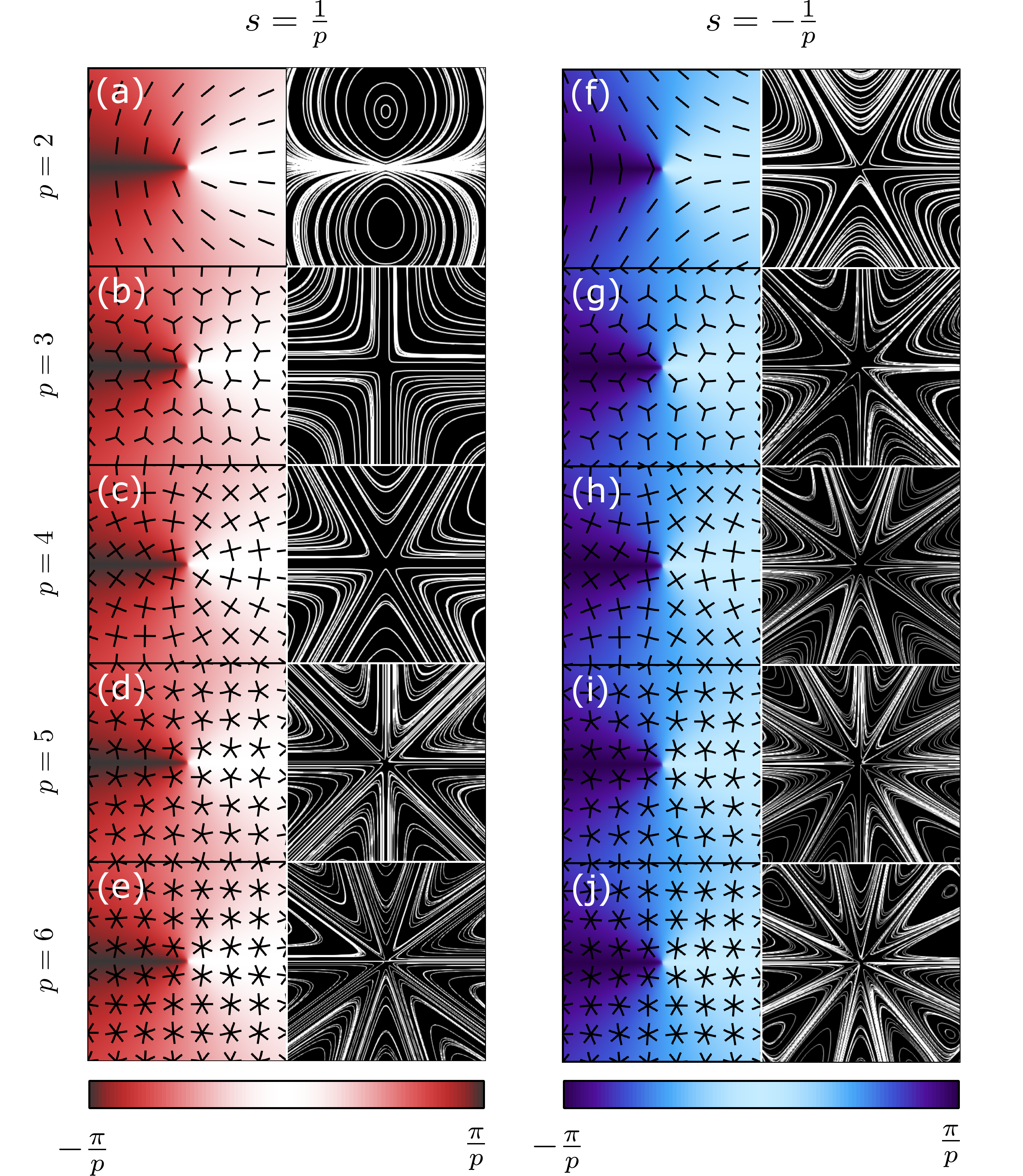}
\caption{\label{fig:1}Elementary $s=1/p$ (red, left column) and $s=-1/p$ (blue, right column) defects and their associated backflow (black and white) for $p=2,\,3\ldots\,6$. Streamlines are obtained from the analytical solutions for the flow field, whose explicit expression is given in Eq.~(S10)~\cite{SI}.}
\end{figure}

We consider an {\em incompressible} $p-$atic liquid crystal, whose microscopic orientation is characterized by the unit vector $\bm{\nu}=\cos\vartheta\,\bm{e}_{x}+\sin\vartheta\,\bm{e}_{y}$ and physical properties are invariant under rotations by $2\pi/p$. For $p=2$, $\bm{\nu}$ is the direction of the rod-like building blocks comprising nematic liquid crystals, for $p=3$, $\bm{\nu}$ is one of the three equivalent directions depicted by the legs of a tri-star, etc, see Fig.~\ref{fig:1}.

At length scales larger than the size of the building blocks, and yet infinitesimal compared to the system size, $p-$atic order can be conveniently described in terms of the tensor order parameter $\boldsymbol{Q}_{p} = Q_{i_{1}i_{2}\ldots\,i_{p}}\bm{e}_{i_{1}} \otimes \bm{e}_{i_{2}} \otimes \ldots \otimes \bm{e}_{i_{p}}$, where $i_{n}= \{x,y\}$ and $n=1,\,2\,\ldots\,p$, constructed upon averaging the $p-$fold tensorial power of the local orientation $\bm{\nu}$~\cite{Giomi:2021a,Giomi:2021b}. The fluid dynamics is in turn governed by the following set of hydrodynamic equations for momentum density $\rho\bm{v}$ and $\bm{Q}_{p}$:
\begin{subequations}\label{eq:hydrodynamics}
\begin{gather}
\rho\,\frac{D\bm{v}}{Dt} = \nabla\cdot\bm{\sigma}\,,\\
\frac{D\bm{Q}_{p}}{Dt} = \Gamma_{p}\bm{H}_{p} + p \big \llbracket \bm{Q}_{p}\cdot\bm{\omega} \big \rrbracket \notag  + \lambda_{p} \big\llbracket \nabla^{\otimes (p-2)}\bm{u} \big\rrbracket \\ 
+ \Bar{\lambda}_p \mbox{tr}(\bm{u}) \bm{Q}_p + \nu_{p} \big \llbracket \nabla^{\otimes (p\,{\rm mod}\,2)} \bm{u}^{\otimes\lfloor p/2 \rfloor} \big \rrbracket\,,
\end{gather}	
\end{subequations}
with $D/Dt=\partial_{t}+\bm{v}\cdot\nabla$ the material derivative, $\rho$ a constant density, $\bm{v}$ the incompressible velocity field ($\nabla \cdot \bm{v} =0$) and $\St$ the total stress tensor. Due to incompressibility, $\mbox{tr}(\bm{u})= \nabla \cdot \bm{v} =0$, and hence term $ \Bar{\lambda}_p \mbox{tr}(\bm{u}) \bm{Q}_p$ in Eq.~(\ref{eq:hydrodynamics}b) vanishes. The tensors $\bm{u}=[\nabla\bm{v}+(\nabla\bm{v})^{\rm T}]/2$ and $\bm{\omega}=[\nabla\bm{v}-(\nabla\bm{v})^{\rm T}]/2$, with ${\rm T}$ indicating transposition, are respectively the strain rate and vorticity fields and entail the coupling between $p-$atic order and flow, with $\lambda_{p}$ and $\nu_{p}$ material constants. The operator $\traceless{\cdots}$ renders its argument traceless, while the dot product indicates the contraction between the last index of $\bm{Q}_{p}$ and the first index of $\bm{\omega}$, $\left( \nabla^{\otimes n} \right)_{i_{1} i_{2} ... i_{n}} = \partial_{i_{1}} \partial_{i_{2}}\ldots\,\partial_{i_{n}}$, whereas $\lfloor \cdots \rfloor$ denotes the floor function; $p\,{\rm mod}\,2=p-2\lfloor p/2 \rfloor$ is zero for even $p$ values and one for odd $p$ values. $\bm{H}_{p}=-\delta F/\delta\bm{Q}_{p}$ is the $p-$atic analog of the molecular tensor, dictating the relaxation dynamics of the order parameter tensor toward the minimum of the orientational free energy $F = \int {\rm d}A\,\left(L_{p}/2\,|\nabla \bm{Q}_{p}|^{2}+A_{p}/2\,|\bm{Q}_{p}|^{2}+B_{p}/4\,|\bm{Q}_{p}|^{4}\right)$, where $|\cdots|^{2}$ is the Euclidean norm and is such that $|\bm{Q}_{p}|^{2} = |\Psi_{p}|^{2}/2$. The constant $L_{p}$ is the order parameter stiffness, while $A_{p}$ and $B_{p}$ are phenomenological constants setting the magnitude of the coarse-grained complex order parameter at equilibrium: $|\Psi_{p}| = |\Psi_{p}^{(0)}| = \sqrt{-2 A_{p}/B_{p}}$, when $H_{i_{1}i_{2}\cdots\,i_{p}}=0$.

We customarily decompose the total stress $\St$ into a static and a dynamic contribution: $\St=\Ss+\Sd$. The static stress tensor is given by $\Ss=-P\mathbb{1}+\Se$, where $P$ is the pressure and $\sigma_{ij}^{({\rm e})} = -L_{p}\partial_{i}\bm{Q}_{p}\odot\partial_{j}\bm{Q}_{p}$ the elastic stress resulting from a static distortion of the $p-$atic orientation; the symbol $\odot$ indicates a contraction of all matching indices of the two operands yielding a tensor whose rank equates the number of unmatched indices (two in this case). We further decompose the dynamic stress into a {\em viscous} or energy dissipating part and a {\em reactive} or energy preserving part: $\Sd=\Sv+\Sr$. The former is given by $\Sv=2\eta\traceless{\bm{u}}$, with $\eta$ the shear viscosity, while the latter takes the form
\begin{equation}
\Sr
=  \lambda_{p}(-1)^{p-1}\nabla^{\otimes p-2}\odot\bm{H}_{p} +
\frac{p}{2}\left(\bm{Q}_{p}\cdot\bm{H}_{p}-\bm{H}_{p}\cdot\bm{Q}_{p}\right)\hspace{-1pt}.
\label{eq:stress_d}
\end{equation}
Both terms describe a departure from the lowest free energy state and, together with the elastic stress $\Se$, can drive backflow (see e.g. Ref.~\cite{Oswald:2005}).  
To investigate the role of backflow in the dynamics of topological defects, we first study isolated disclinations of strength $s=\pm 1/p,\,\pm 2/p\ldots$ at the origin of an unbounded domain and take the phase of the coarse-grained complex order parameter $\theta=s\phi+\theta_{0}$, with $\phi=\arctan(y/x)$. The constant angle $\theta_{0}$ determines the overall orientation of the defect~\cite{Vromans:2016} and can be set to zero without loss of generality. $|\Psi_{p}|$ is assumed uniformly equal to its equilibrium value $|\Psi_{p}^{(0)}|$ outside the core of the defects $-$ i.e. for $|\bm{r}|>a$ with $a\sim\mathcal{O}(\sqrt{L_{p}/|A_{p}|})$ the defect core radius $-$ and zero inside it. Following Refs.~\cite{Giomi:2013,Giomi:2014}, we assume flow alignment effects to be negligible and compute the stationary solution of Eq.~(\ref{eq:hydrodynamics}a). Interestingly, of the three backflow-driving terms in the total stress tensor, only the first term in Eq.~\eqref{eq:stress_d} contributes to the flow surrounding the defect; the flow alignment stress. 
By contrast, the second term $-$ the antisymmetric stress $-$ which originates from the corotational derivative of the tensor order parameter, is proportional to $\nabla^{2}\theta$ when $|\Psi_{p}|={\rm const}$ (see e.g. Ref.~\cite{Giomi:2021b}), and hence vanishes identically. The elastic stress, on the other hand, yields the isotropic force density $\nabla\cdot\Se=(ps|\Psi_{p}|)^{2}L_{p}\,\bm{r}/(2|\bm{r}|^{4})$, which, in turn, leads to a local pressure variation, $P\to P-(ps|\Psi_{p}|)^{2}L_{p}/(2|\bm{r}|^{2})$, without influencing the flow. 

To obtain the backflow sourced by isolated defects, we convolute the two-dimensional Oseen-Green tensor~\cite{SI} with the body force $\bm{f}=\nabla\cdot\Sr=c_{p}/|\bm{r}|^{p+1}[\cos(n\phi)\,\bm{e}_{x}+\sin(n\phi)\,\bm{e}_{y}]$, where $c_{p}$ and $n$ are given by
\begin{subequations}\label{eq:backflow_force}
\begin{gather}
c_{p} = \frac{ (-1)^{p}(ps)^{2}\lambda_{p}L_{p}}{2^{p/2}}~ \prod^{p-1}_{k=1}~ \left[ps-2(p-k)\right]\;,\\
n = p(s-1)+1\;.
\end{gather}	
\end{subequations}
The resulting flow field surrounding the defects is given in Eq.~(S10)~\cite{SI}. Our solutions for the flow field hold for {\em arbitrary} $p$ values, thus include nematics.

We plot the backflow of the thermodynamically stable defects $s=\pm 1/p$ for $p=2,\,3\ldots\,6$ in Fig.~\ref{fig:1}. The backflow of defects of \textit{all} $s$ inherits the $(|n|+1)-$fold rotational symmetry of the driving force $\bm{f}$, with $n$ found in Eq.~(\ref{eq:backflow_force}b). Thus, $s=1/2$ disclinations in nematics, source a typical Stokeslet-like flow consisting of two counter-rotating vortices meeting along the defect's longitudinal direction (i.e. the $x-$direction in this case), yielding a net momentum current, whose effect is to propel the defect forward. By contrast, $s=-1/2$ give rise to a $3-$fold symmetric flow consisting of six vortices with alternating positive and negative vorticity. Similarly, $s=1/3$ ($s=-1/3$) disclinations in {\em triatics} drive a $2-$fold ($4-$fold) symmetric flow, etc.  Due to this rotational symmetry, these flows stir the fluid around a defect and trap the core at the central stagnation point, for all $p-$atic defects with $n\neq0$.

\begin{figure}[b!]
\centering
  \includegraphics[width=\linewidth]{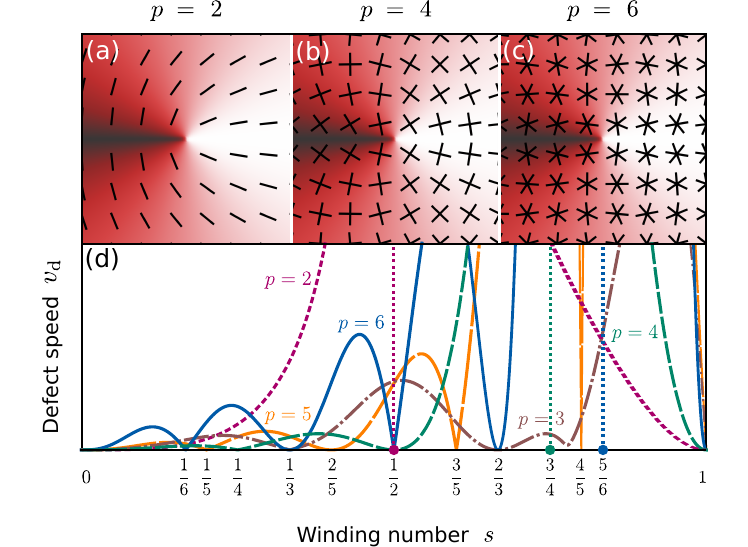}
\caption{\label{fig:2}(a) Motile defects with $s=1/2$, $3/4$ and $5/6$. (b) Speed of the defect core for $0\le s\le 1$. Notice that speed is zero for all defects of strength $s=1/p$ apart from nematics, where $s=1/2$.}
\end{figure}

For $n=0$, on the other hand, defects propel under the effect of their associated backflow. This leads to self-propulsion condition that is valid of {\em all} $p$ values: i.e. $s=(p-1)/p$ with $p$ even (Fig.~\ref{fig:2}a-c). For odd $p$ values, on the other hand, $c_{p}=0$ when $n=0$ and the driving force $\bm{f}$ vanishes (Fig.~\ref{fig:2}d). The speed of self-propelled defects is readily found upon integrating the velocity field $\bm{v}$, Eq.~(S10) in Ref.~\cite{SI}, along the defect core. This gives  
\begin{multline}\label{eq:defect_speed}
v_{\rm d} = \mu_{p}~R^{1-p} \\ \left\{p+1+\left(\frac{a}{R}\right)^{1-p}\left[\frac{3p-1}{2}+(p-1)\log\frac{a}{R}\right]\right\}\;,
\end{multline}
where $\mu_{p}=(-1)^{p+1}\pi\lambda_{p}L_{p}/(2^{p/2+1}\eta)\prod_{k=1}^{p-1}(k-p-1)$. With exception of $p=2$, however, none of these self-propelled defects feature an elementary winding number $s=\pm 1/p$, and when allowed to evolve rapidly split into $p-1$ elementary defects.
We note that the fact that self-propulsion is solely driven by the flow alignment stress has not been previously identified even for nematics, where self-propulsion has been a subject of thorough investigation~\cite{toth2002,toth2003hydrodynamics,kats2002disclination,svenvsek2002hydrodynamics,PhysRevResearch.2.013080,kos2020field}.
In fact, the proportionality between the speed of isolated $s=1/2$ disclinations and the flow alignment parameter brings to light an exciting opportunity for estimating the flow alignment parameter $\lambda_{2}$ $-$ a notoriously elusive material parameter in liquid crystals (see e.g. Ref.~\cite{PhysRevE.70.011701}) $-$ from measurements of the self-propulsion speed of elementary nematic defects.

To obtain an exhaustive understanding of how backflow affects defect dynamics, we must examine how it affects their interactions.
Hence, we study the annihilation dynamics of neutral elementary defect pairs, $s=\pm 1/p$. In the absence of hydrodynamic effects, two-dimensional disclinations of opposite strength are known to attract via a Coulomb-like force and eventually annihilate \cite{de1993physics}. In nematics, T\'oth {\em et al}. showed that hydrodynamics affects this process in a two-fold way~\cite{toth2002}: first, advection by backflow causes defects to move faster, thereby speeding up their annihilation dynamics. Subsequently, the different configuration of the velocity field surrounding positive and negative defects introduces an asymmetry in the annihilation trajectory, which is then no longer symmetric about the mid-plane separating the defects at $t=0$. Although tempting to explain this phenomena in light of the aforementioned passive self-propulsion, in what follows we uncover that, surprisingly, both effects are primarily fueled by the antisymmetric part of the dynamic stress, i.e. the second term of Eq.~\eqref{eq:stress_d}.  
\begin{figure}[t!]
\centering
  \includegraphics[width=0.49\textwidth]{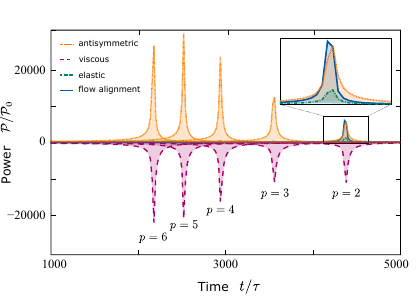}
\caption{\label{fig:3}Power $\mathcal{P}$ delivered by each and every contribution to the total stress, for $p=2,\,3\ldots\,6$. The dominant contribution to the backflow (yellow tones) is sourced by the antisymmetric component of the stress and dissipated by the viscous stress (magenta tones). Inset: $\mathcal{P}$ delivered by all components of the stress close to the annihilation time for nematic defects. Only in this case, $\mathcal{P}$ arising from the first term in Eq.~\eqref{eq:stress_d} (blue) is comparable in magnitude to the antisymmetric stress's dominant contribution; $\mathcal{P}$ has been rescaled by $\mathcal{P}_{0}=L_{p}/\tau$, with $\tau=a^{2}/(\Gamma_{p}L_{p})$.}
\end{figure}

To demonstrate this, we meticulously analyze the power $\mathcal{P} =\int {\rm d}A\,[\nabla\cdot\bm{\sigma}]\cdot\bm{v}$ delivered by each and every contribution to the total stress, before, during, and after, an annihilation event (Fig.~\ref{fig:3})~\cite{carenza2020_bif,carenza2020}. Data is generated by numerical integration of Eqs.~\eqref{eq:hydrodynamics} on a periodic square domain, with initial configuration consisting of a neutral pair of elementary $p-$atic defects~\cite{SI,Carenza2019}. Our analysis reveals that for all $p$ values, the annihilation dynamics is dominated by the antisymmetric part of the stress tensor (yellow tones). This converts the energy stored in the distorted configuration of the $p-$atic director into kinetic energy, which is in turn dissipated by viscous stresses (magenta tones). By contrast, stresses originating from flow alignment, sourcing the propulsion of isolated defects, contribute to the annihilation dynamics only for $p=2$ (Fig.~\ref{fig:3} inset), i.e. when the second and third terms in Eq.~\eqref{eq:stress_d} have the same differential order. 

\begin{figure}[t!]
\centering
  \includegraphics[width=\columnwidth]{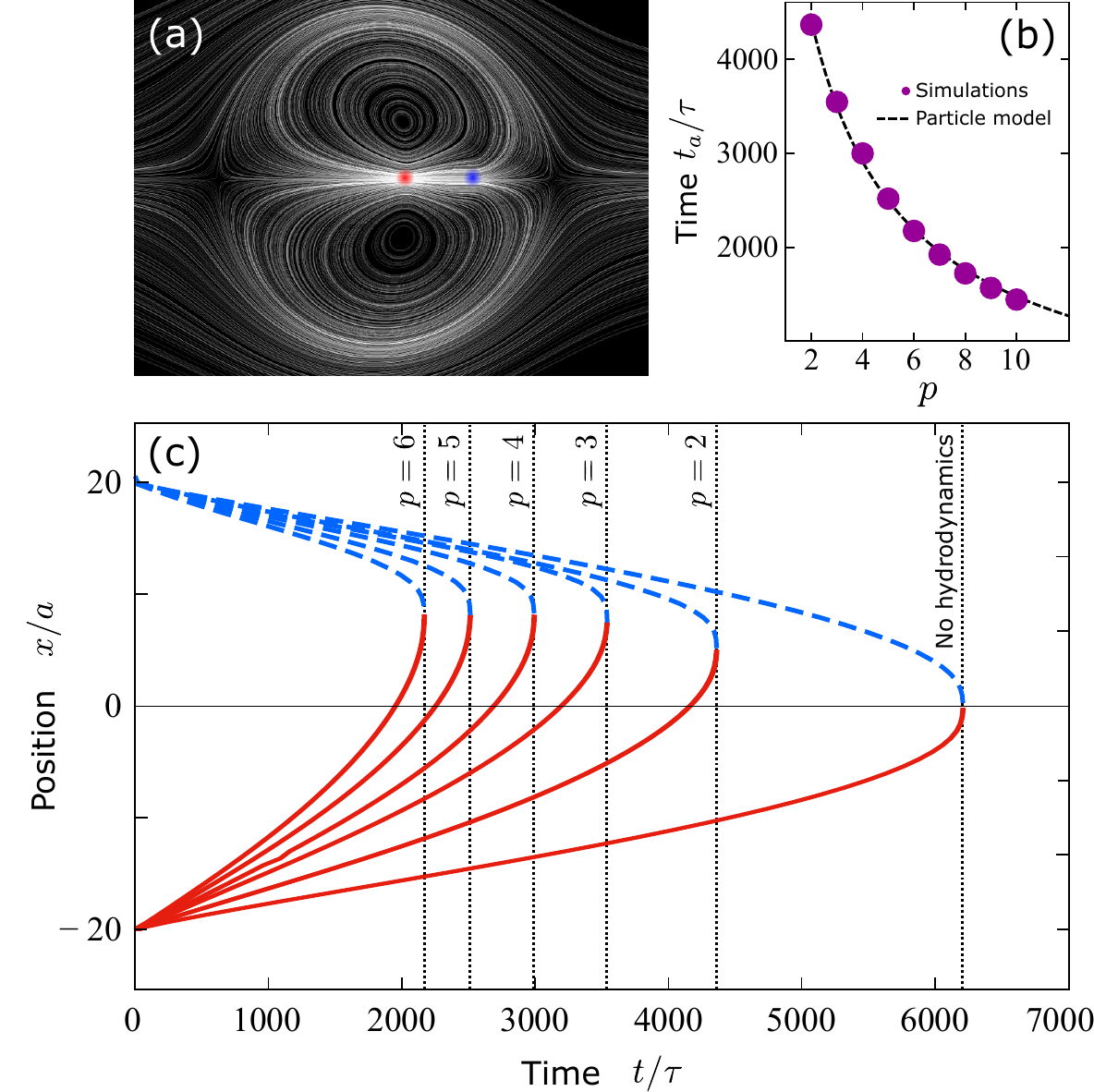}
\caption{\label{fig:4} (a) Flow field surrounding a $\pm 1/p$ pair obtained from numerical integration of Eqs.~\eqref{eq:hydrodynamics}. Red (left) and blue (right) dots denote the defects' position.
(b) The annihilation time $t_a$ decreases with $p$, in a decreasing rate. (c) Trajectories of annihilating defects in time. The asymmetry in the trajectories increases until it reaches its maximum value. The continuous and dashed lines denote the trajectories of the positive and negative defects, respectively.}
\end{figure}

Beyond revealing the origin of the hydrodynamic enhancement of pair annihilation, we illustrate yet another striking result in Fig.~\ref{fig:4}: annihilation occurs more rapidly as $p$ increases. To elucidate this phenomenon, we focus on trajectories of annihilating defects, with and without backflow (Fig.~\ref{fig:4}a). In the presence of hydrodynamics the positive defect moves faster towards the positive $x-$direction and annihilation occurs in the half-plane initially occupied by the negative defect  (Fig.~\ref{fig:4}b,c). Most surprisingly, this phenomenon becomes more prominent for hexatics, consistently with our observation that annihilation occurs more rapidly as $p$ increases. Comparing the annihilation trajectories of all $p$ values in the range $p=2,\,3\ldots\,6$, however, we find that, even though the annihilation time $t_{\rm a}$ decreases with $p$, it does so at a decreasing rate  (Fig.~\ref{fig:4}c). This behavior can be rationalized from simple force balance. Calling $x_{\pm}$ the positions of the defects along the $x-$axis, this implies
\begin{equation}\label{particle model}
\varsigma\left(\frac{{\rm d}x_{\pm}}{{\rm d}t} - v_{\pm}\right) = - \frac{\mathcal{D}s^{2}}{x_{\pm} - x_{\mp}}\;.
\end{equation}
The left-hand side of the equation denotes the effective drag force experienced by the defects, with $v_{\pm} \sim p$ the speed of the propelling backflow and $\varsigma \sim s^{2}$ a drag coefficient~\cite{Denniston:1996}. The right-hand side arises from the elastic Coulomb attraction between defects, with $\mathcal{D} \sim \Gamma_{p}L_{p}$ a rotational diffusion coefficient independent of $p$. To compute the annihilation time $t_{\rm a}$ we solve Eq.~\eqref{particle model} imposing  $\Delta v = v_{+}-v_{-}\ge 0$ and $x_{+}(t_{\rm a})=x_{-}(t_{\rm a})$, yielding
\begin{equation}\label{annihilation time}
t_{\rm a} =  \frac{|\Delta x(0)|}{\Delta v}-\frac{2\mathcal{D}s^{2}/\varsigma}{\Delta v^{2}}\log\left[1+\frac{|\Delta x(0)|}{2\mathcal{D}s^{2}/\varsigma}\,\Delta v\right]\;,
\end{equation}
where $\Delta x(0) = x_{+}(0)- x_{-}(0)$. In the absence of backflow, $\Delta v=0$ and Eq.~\eqref{annihilation time} reduces to $t_{\rm a} = |\Delta x(0)|^{2}/(4\mathcal{D}s^{2}/\varsigma)$. For finite $\Delta v$ values, on the other hand, $t_{\rm a}$ decreases monotonically with $\Delta v$ and approaches $t_{\rm a}\approx |\Delta x(0)|/\Delta v$ for large $\Delta v$ values. In turn, $\Delta v$ increases with $p$ for small $p$ values~\cite{SI}, but vanishes for $p\to\infty$ when isotropy is restored at the macroscopic scale and the defects themselves disappear: i.e. $s=\pm 1/p\to 0$.

In conclusion, we have demonstrated that backflow profoundly affects $p-$atic defect dynamics in two principal ways, as well as identified the origins of its effects. First, we showed that backflow fuels a \textit{generic} self-propulsion mechanism for of all defects with winding number $s=(p-1)/p$, which is however thermodynamically stable only in nematics. Second, we discovered that backflow \textit{always} accelerates the dynamics of neutral elementary defect pairs $s=\pm 1/p$, and, contrary to expectations, becomes \textit{increasingly} more relevant as $p$ increases. The latter is readily amenable to experimental scrutiny, for instance in suspensions of lithographically printed colloidal polygons~\cite{Wang2018}. Furthermore, for space-filling polygons ($p > 3$), faster pair annihilation can enhance the coarsening dynamics during crystallization, with important potential applications to fabrication of ordered monolayers for bio-medicine, optics etc.~\cite{Denkov1993}. 
Finally, our work paves the way towards understanding cell intercalation and other remodelling events \cite{saw2017,kawaguchi2017,Balasubramaniam2021,Hoffmann2022,Streichan2018,duclos2017} in epithelial layers, where small-scale hexatic order ($p = 6$)
has been recently discovered~\cite{Collado2022a,Collado2022b}.

\acknowledgements

This work is supported by the ERC-CoG grant HexaTissue and by Netherlands Organization for Scientific Research (NWO/OCW). Part of this work was carried out on the Dutch national e-infrastructure with the support of SURF through the Grant 2021.028 for computational time. The authors acknowledge Ludwig Hoffmann for fruitful discussions.

%\bibliography{bib.bib}
\bibliographystyle{apsrev4-2}

\end{document}